\documentclass[journal]{IEEEtran}
\ifCLASSINFOpdf
\else
\fi
\usepackage{hyperref}
\hypersetup{hidelinks,
	colorlinks=true,
	allcolors=black,
	pdfstartview=Fit,
	breaklinks=true}
\usepackage{float}
\usepackage{booktabs}
\usepackage{multirow}
\usepackage{algorithm}
\usepackage{algorithmic}
\usepackage{xcolor, soul}
\usepackage{verbatim}
\usepackage{graphicx}
\usepackage{array}
\usepackage{amsmath}
\usepackage{subfig}
\usepackage{threeparttable}   

\usepackage{tikz,xcolor,hyperref}
\definecolor{lime}{HTML}{A6CE39}
\DeclareRobustCommand{\orcidicon}{%
    \begin{tikzpicture}
    \draw[lime, fill=lime] (0,0) 
    circle [radius=0.16] 
    node[white] {{\fontfamily{qag}\selectfont \tiny ID}};    \draw[white, fill=white] (-0.0625,0.095) 
    circle [radius=0.007];    \end{tikzpicture}
    \hspace{-2mm}}
\foreach \x in {A, ..., Z}{%
    \expandafter\xdef\csname orcid\x\endcsname{\noexpand\href{https://orcid.org/\csname orcidauthor\x\endcsname}{\noexpand\orcidicon}}
    }

\usepackage{color}
\usepackage{hyperref}
\hypersetup{
    colorlinks=true,
    linkcolor=blue,
    filecolor=blue,      
    urlcolor=black,
    citecolor=blue,
}

\begin{document}

\title{DVS-RG: Differential Variable Speed Limits Control using Deep Reinforcement Learning with Graph State Representation }



\author{Jingwen~Yang\orcidA{},~\IEEEmembership{Graduate Student Member,~IEEE,}
        Ping~Wang\orcidB{},~\IEEEmembership{Member,~IEEE,} \\
        Fatemeh~Golpayegani\orcidC{},~\IEEEmembership{Senior~Member,~IEEE,}
        and~Shen~Wang\orcidD{},~\IEEEmembership{Senior~Member,~IEEE}
\thanks{Manuscript received; revised.}
\thanks{Jingwen Yang is with the School of Electronic and Control Engineering, Chang’an University, Xi’an 710054, Shaanxi, China. (e-mail:2020032001@chd.edu.cn.)}
\thanks{Ping Wang is with the School of Intelligent Systems Engineering, Sun Yat-sen University, Shenzhen, Guangdong 510006, China. (e-mail:wangp358@mail.sysu.edu.cn)}
\thanks{Fatemeh Golpayegani and Shen Wang are with the School of Computer Science, University College Dublin, Dublin 4, D04 V1W8 Ireland (e-mail: Fatemeh.golpayegani@ucd.ie; shen.wang@ucd.ie).}
}

\markboth{Journal of \LaTeX\ Class Files,~Vol.~, No.~}%
{Shell \MakeLowercase{\textit{et al.}}: Bare Demo of IEEEtran.cls for IEEE Journals}
\maketitle

\begin{abstract}
Variable speed limit (VSL) control is an established yet challenging problem to improve freeway traffic mobility and alleviate bottlenecks by customizing speed limits at proper locations based on traffic conditions. Recent advances in deep reinforcement learning (DRL) have shown promising results in solving VSL control problems by interacting with sophisticated environments. However, the modeling of these methods ignores the inherent graph structure of the traffic state which can be a key factor for more efficient VSL control. Graph structure can not only capture the static spatial feature but also the dynamic temporal features of traffic.
Therefore, we propose the DVS-RG: DRL-based differential variable speed limit controller with graph state representation. DVS-RG provides distinct speed limits per lane in different locations dynamically. The road network topology and traffic information(e.g., occupancy, speed) are integrated as the state space of DVS-RG so that the spatial features can be learned. The normalization reward which combines efficiency and safety is used to train the VSL controller to avoid excessive inefficiencies or low safety. The results obtained from the simulation study on SUMO show that DRL-RG achieves higher traffic efficiency (the average waiting time reduced to 68.44\%) and improves the safety measures (the number of potential collision reduced by 15.93\% )  compared to state-of-the-art DRL methods. 
\end{abstract}


\IEEEpeerreviewmaketitle

\section{Introduction}

\subsection{Background}
\begin{figure}[!ht]
	\centerline{\includegraphics[width=0.75\columnwidth]{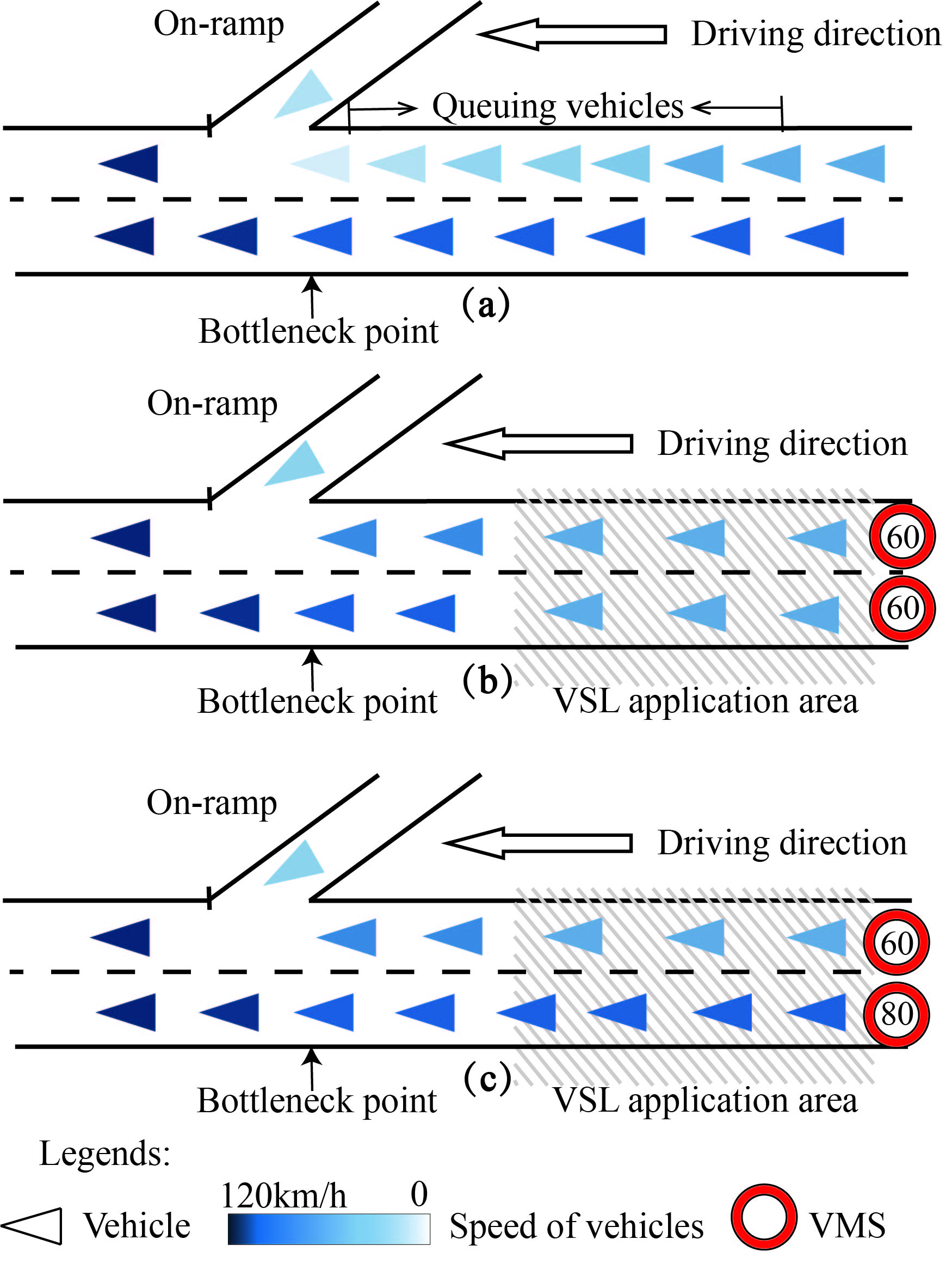}}
	\caption{(a) The traffic bottleneck performance under NO-VSL; (b) The traffic bottleneck performance under VSL; (c) The traffic bottleneck performance under DVSL.}
\label{fig.RL_VSL}
\end{figure}

Traffic bottlenecks disrupt the vehicular traffic, and result in increased vehicular delays, driving stress, environmental pollution, and reduced traffic safety and efficiency\cite{chen2014variable,wang2022integrated,wu2020differential}.
Once a traffic bottleneck occurs, vehicles will quickly accumulate in the upstream sections, which commonly happens at entrance ramps and sections where lanes are reduced, as shown in Fig. \ref{fig.RL_VSL}a.
Variable speed limits(VSLs) can limit the traffic flow into the bottleneck section and are an effective way to eliminate traffic bottlenecks among mainstream traffic flow control \cite{wang2021freeway}. The VSL can be applied upstream of the bottleneck point to harmonise the speed across vehicles to regulate the mainstream arriving flow, as shown in Fig. \ref{fig.RL_VSL}b. The adaptive VSL can change posted speed limits by displays on overhead or roadside variable message signs (VMS) in response to the development of the prevailing traffic information. VSLs systems can improve throughput and traffic safety and ensure stable traffic flow affected by congestion through balanced speed and utilization of lanes. 
Moreover, the differential variable speed limits (DVSL) method has sparked widespread interest among researchers, which dynamically sets distinct speed limits for each lane, and is capable of offering precise and timely lane-level VSL instructions, as illustrated in Fig. \ref{fig.RL_VSL}c. DVSLs have exhibited advantages in congestion alleviation, as well as accident and emission reductions \cite{wu2020differential,lu2023td3lvsl} .  

\subsection{Related work}
The traditional VSL method can be divided into two types: reactive and proactive.
The reactive VSL control includes open-loop \cite{chen2014variable} and ruled-based \cite{yuan2022selection} techniques. As previously mentioned, realtime VSL decisions changed based on preselected thresholds of traffic flow, occupancy, or mean speed \cite{khondaker2015variable}. Although the logic of the above methods is simple and easy to deploy \cite{weikl2013traffic}, its major limitation is time lag. 
The proactive VSL method have been developed to address the limitations of the reactive counterparts, which can dynamically adjust VSL decisions based on the feedback of the system \cite{jin2015control,du2019variable,karafyllis2019feedback,carlson2011local,muller2015microsimulation}. The most representative is the model predictive control (MPC), which was used to prevent traffic jams by predicting the traffic flow \cite{mao2022variable,guo2020integrated,han2021lineara}.
The core idea is traffic breakdowns are anticipated before they even occur, and remedial VSL strategies are implemented and resolve shockwaves before traffic reaches breakdown. However, building an accurate and reliable traffic flow prediction model is challenging, because the microscopic drivers’ behavior, such as sudden deceleration, merging, or lane changing, resulting in uneven headways, cannot be completely reflected in the model. 

Deep reinforcement learning (DRL) is a model-free and data-driven approach that is used as an alternative to model-based approaches \cite{ye2020modelfree} in the context of traffic panning and control \cite{guo2023cotv,yang2021automaticb}. DRL consists of an agent gradually learning the optimal control policy by utilizing experiences acquired from repeated interactions with the environment,  without predefined human rules or models in a complex environment \cite{silver2016mastering,bai2022hybrid}. As a result, a well-trained DRL agent can, theoretically, make predictions on system evolution and achieve a proactive control scheme. DRL-based VSL has been also proposed, which can be divided into 3 categories\cite{kusic2020overview}:

In \textbf{value-based} methods, the Q-learning (QL) algorithm is used in early research \cite{walraven2016traffica,li2017reinforcement,wang2019new,han2022new}. The common problem is Q table used to record experiences is limits space. To address the curse of dimensionality for large-scale traffic control, Walraven et al. \cite{walraven2016traffica} proposed function approximation techniques. Wang et al. \cite{wang2019new} proposed a multi-agent system with a distributed QL algorithm to tackle the cooperative control problem in continuous traffic state space. Furthermore,  Kisic et al. \cite{kusic2021spatialtemporalc} proposed a deep Q-learning-based (DQL) VSL algorithm including a customized learning process and a complex reward function consisting of three separate objectives. The value-based methods including QL or DQL rely on finding an action which maximizes the action-value function, however, they discretize the action domain for the applications with continuous action variables. The discretization of the action domain may lead to the curse of the dimensionality issue since the number of total actions increases exponentially with the number of action types. Moreover, the discretization of action space may cause information loss and lead to sub-optimal solutions. This makes it intractable to apply the value-based method to applications with high-dimension and continuous action space. 

The second category includes the  \textbf{policy-based} methods. Peng et al. \cite{peng2022combined} developed a VSL controller based on Proximal Policy Optimization(PPO) and the training curve shows that the performance of the controller has been improved via the PPO algorithm with increasing reward and decreasing loss. Wang et al. \cite{wang2022integrated} developed a centralized traffic control system that can coordinate multiple ramp metering and VSL traffic controllers on freeways to minimize the total travel time. The results show that the deep deterministic policy gradient (DDPG) and twin delayed deep deterministic policy gradient (TD3) outperform other control methods and the reward curve of TD3 is more stable than DDPG. Although the training curves of both methods have an upward trend, the fluctuations are obvious.

In \textbf{actor and critic} methods, Wu et al. \cite{wu2020differential} developed a DVSL controller based on DDPG, in which dynamic and distinct speed limits per lane can be imposed. Different reward signals are used to train the DVSL controller, and a comparison between these reward signals is conducted. Their experiments have shown that their proposed method can reduce congestion, as well as accidents and emissions in a simulation study. However, most of their training reward curves reported in the paper do not seem to converge.

Above all, there are some proposed customizing neural network methods for VSL(DVSL) controllers \cite{wang2022integrated,wu2020differential}, which may have hyperparameter sensitivity, training instability, etc. The massive input traffic data and the evolving behaviours of agents (controllers) in the global environment cause nonstationarity and instability during training, which is a common phenomenon \cite{wu2020differential,devailly2022igrla}. What‘s more,  VSL traffic flow with temporal and spatial is necessary to provide valuable information for the VSL method which is ignored by most studies. As the author knows, there is no literature that integrates topological relationships and traffic conditions information into DRL-based VSL(DVSL).


\subsection{Scope and Contributions}


To address the challenges mentioned above, we propose DVS-RG: DVSL control combined with DRL and graph state representation to improve the traffic efficency and safety during the merging area, which is is also open-sourced here\footnote{\url{https://github.com/jingwenyanga/DVS-RG/}}.
A graph is a data structure with great expressive power to model a set of objects (lanes) and their relationships and is already successfully applied in transportation systems\cite{devailly2022igrla,chen2021graph,yoon2021transferable}.
Considering the decisions of DVSL relate not only information pertaining to traffic information in its proximity (temporal information)  but also location information pertaining to coming from downstream or upstream (spatial information),  a directed graph including traffic flow information and segment network topology is proposed.
A directed graph passes with non-linear neighbourhood transformation and then is followed by aggregation to yield a fused representation of spatial embedding to form graph state representation that serves as the state space of DRL.
The state space, which involves traffic information and information dissemination based on the spatial relationship between DVSL lanes and their surroundings, will benefit the DVSL agent in learning policy.
While, the proposed DRL-DVSL framework can be tested in the open-source algorithm , which can avoid the drawbacks of customizing neural networks
The contribution of this paper can be summarized as follows:
\begin{enumerate}
      \item A novel state space design is proposed where we combine the traffic information and the graph representation of the road network. This novel design allows the DVSL agent to capture finer-grained traffic information and its dissemination, which has benefits for training the DVSL agent;
    
    \item Reward normalization is introduced in the design of the reward function, which can filter out some lower efficiency and high collision situations. The normalization layer is introduced to merge different traffic data to speed up training.
    \item Experiments validate the effectiveness of the proposed method compared with state-of-the-art methods. The method has significant improvement in efficiency and safety, up to 68.44\% and 15.93\%, respectively.
\end{enumerate}

\subsection{Paper Organization}

The remainder of this paper is organized as follows. Section \ref{s2} introduces the system design and detail architecture. Section \ref{s3} details the scenario and experiments implemented. Section \ref{s4} presents the results and analysis to demonstrate the value of the proposed method. Finally, Section \ref{s5} discusses conclusions and future extensions of this work.

\section{System Overview}\label{s2}

\subsection{System Design Goals}

The goal of the proposed model is to optimize the freeway bottleneck traffic via DRL-based differential VSL controllers. Optimising traffic efficiency and  safety are the main objectives considered in the proposed control system.

\begin{itemize}
    \item \textbf{High safety}: Safety is the first consideration in traffic control system design. Time-to-collision (TTC) index has been extensively utilized to evaluate rear-end collision risks\cite{lu2023td3lvsl}. TTC is defined as the time remaining to a potential collision if the interacting road users’ speed and direction remain unchanged. To identify  dangerous situations, a critical threshold must be determined for TTC. When the values are over the threshold,  the likelihood of traffic accidents increases. The TTC of each vehicle is calculated and if the TTC is less than the determined threshold, the vehicle will potentially collide. Therefore, the target of the proposed control system regarding safety is to have fewer vehicles with the TTC exceeding the threshold defined. 
    
    \item \textbf{High efficiency}: Average Waiting time (AWT) is an important measure that not only reflects traffic efficiency but also reflects the freeway level-of-service scale. Total stopped time (TST)  may aggravate the phenomenon of stop-and-go shockwaves to some extent, which causes low-efficiency and high-safety. Throughput serves as the evaluation index of traffic efficiency, which represents the number of vehicles that pass through the study site in a time interval. In this paper we are focusing on Bottleneck throughput(BT). 
\end{itemize}

To address the two main objectives of this paper, a control system is designed to decrease the AWT, TST and increase BT  to untimately improve efficiency.  

\subsection{System Architecture}

\begin{figure*}[!ht]
	\centerline{\includegraphics[width=1.6\columnwidth]{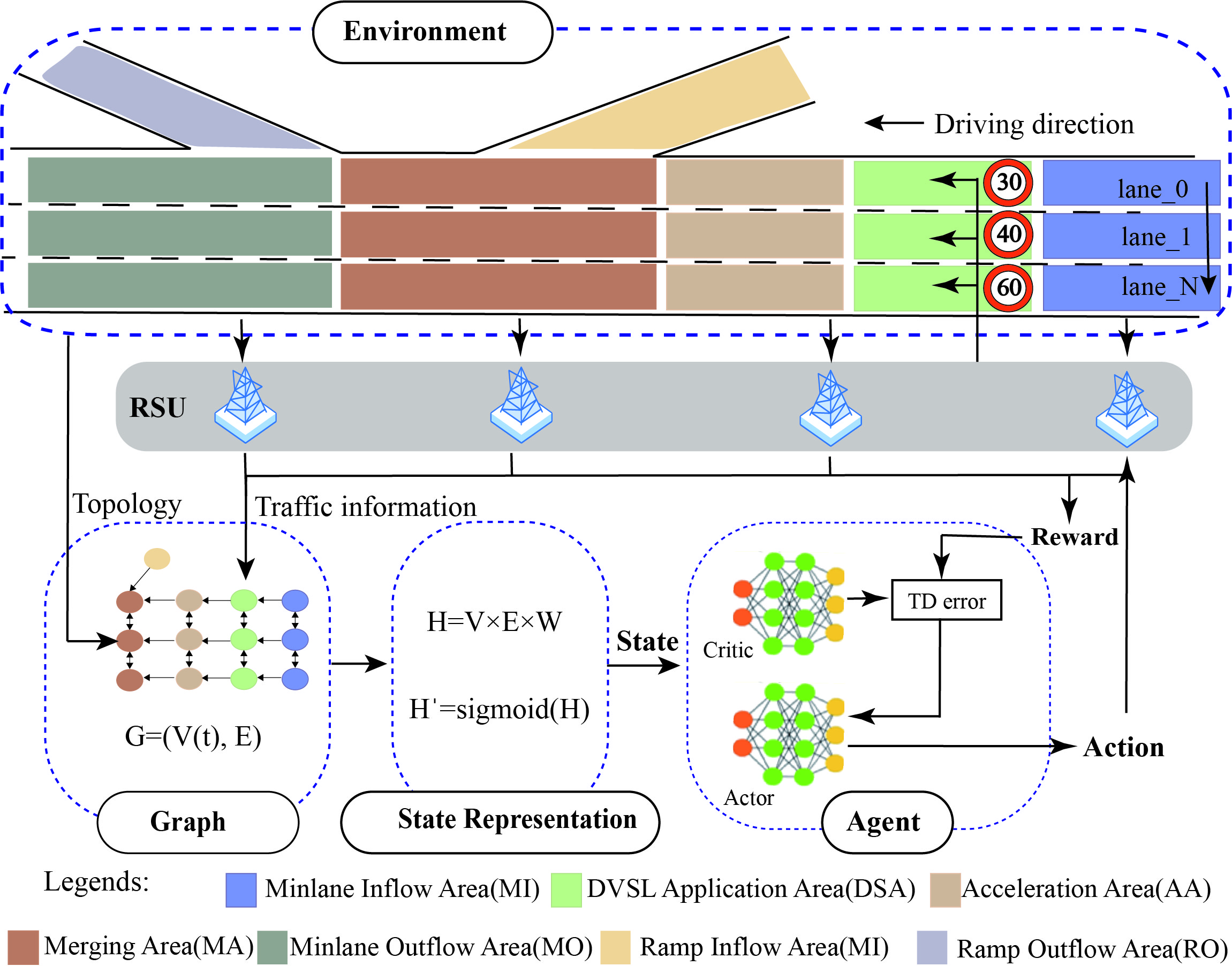}}
	\caption{The actor-critic architecture for DVSL with graph state representation }\label{fig.fam}
\end{figure*}

A traffic control area can be divided into a series of segments, and the segment length should satisfy $\Delta L \geq v_{seg} \Delta t$ according to the Courant-Friedrichs-Lewy (CFL) condition, where $x$ is the segment length, $v_{seg}$ is the average speed of the segment, and $\Delta t$ is the time step. It notices that This work assumes that all the vehicles are connected and the  V2X communication is  assumed to be perfect without packet loss and no latency. Some studies show imposing VSL control some distance upstream of a bottleneck to starve the inflow to the bottleneck and dissipate the queue \cite{chen2014variable,wang2022integrated,muller2015microsimulation,li2017reinforcement}. The acceleration area allows for vehicles that exit from the controlled congestion to accelerate and traverse the bottleneck area with the critical speed. The maximum throughput is achieved at the bottleneck and capacity drop is avoided\cite{wang2021freeway}. Therefore, five areas should be focused for DVSL control: the mainlane inflow area (\textbf{MI}), DVSL application area (\textbf{DSA}), acceleration area(\textbf{AA}) , on-ramp inflow area (\textbf{RI}) and merging area (\textbf{MA}).  Except \textbf{MA}, the length of each area is generally $\Delta L$.
The length of \textbf{MA} is changed according to different scene. The length of \textbf{MA} is generally the distance from the on-ramp point to the off-ramp point. When the distance between the on-ramp point and the off-ramp is bigger than $\Delta L$ or there is no off-ramp, the \textbf{MA} distance equals $\Delta L$ from the on-ramp point(bottleneck point). 
The downstream of the bottleneck point includes the two following areas of interest: mainlane outflow area (\textbf{MO}) and ramp outflow area (\textbf{RO}), as is shown in the upper part of Fig \ref{fig.fam}.

%

The actor-critic is used in this paper for the DVSL controller and is a hybrid architecture combining value-based and policy-based methods that help to stabilize the training by reducing the variance: The actor generates an action for the DVSL strategy, and the critic is utilized to evaluate DVSL strategy.
Policy-based DRL is effective in high dimensional and stochastic continuous action spaces, and learning stochastic policies, and the value-based DRL excels in sample efficiency and stability. The actor-critic architecture integrates the advantages above mentioned.  The algorithm architecture along with the graph state representation used for DVSL is shown in the lower part of Fig \ref{fig.fam}. 
The DVSL controller outputs a set of speed limits for each lane to improve traffic efficiency and safety.
The design of action, state, and reward for the DVSL controller is given as follows:

\subsubsection{\textbf{Action}}
The action set includes several values of the speed limit in continuous space. The action $a$ shown in Eq. \ref{equ:action}

\begin{equation}
a = [a_{1}^{vsl},a_{2}^{vsl},\dots,a^{vsl}_{N_c}]
\label{equ:action}
\end{equation}

\begin{equation}
a^{vsl} = v^{vsl}_{min}+v_m \times u
\end{equation}

\begin{equation}
v^{vsl}_{min} \leq a_{vsl} \leq v^{vsl}_{max}
\label{equ:state_s}
\end{equation}

\noindent where $N_c$ is the number of lanes in the \textbf{DSA}; $u$ is the output of the agent, $u \in [0,1]$; $v_m$ is a coefficient; 
The $v_{vsl}$ is the value of the speed limit of each lane and is required to be greater than the minimum speed limit  $v^{vsl}_{min}$ and not greater than the maximum speed limit $v^{vsl}_{max}$.

\subsubsection{\textbf{Reward}} 

The reward is calculated at every time step $t$. The efficiency and safety of the system are considered the two main factors during our reward design process.

For efficiency, we consider the average speed of the \textbf{MA} as the most important index. The average speed is normalized to facilitate the DRL training process. The calculation of this index is as follows: 

\begin{equation}
v_{MA}
=\begin{cases}
0, & \text{if $\exists$ $v_{MA,i}$ $<$ $ v_{c,min}$}  \\
\frac{\sum_{i=1}^{N_{MA}}v_{MA,i}-v_{c,min}}{v_{max}-v_{c,min}}, & \text{otherwise}
\end{cases}
\end{equation}

\begin{equation}
\overline{v}_{MA}= \frac{1}{T_c}\sum_{t=1}^{T_c}v_{MA}(t)
\end{equation}

\noindent where $v_{MA}$ reflect the average speed of MI and is calculated by the speed of each lane; if the speed of any MI lane is less than the customized critical minimize speed value $v_{c, min}$, the $v_{MA}$ is equal to its average; otherwise, the $v_{MA}$ is equals 0; $v_{max}$ is the maximum speed allowed by the road; $v_{MA, i}$ is the speed of lane $i$ in MA; $T_c$ is the time of a control horizon; $\overline{v}_{MA}$ is the average speed of MA during a control horizon

For safety, TTC is an effective index for identifying traffic conflict and evaluating the safety of the traffic flow, estimating the time required for a car to hit its preceding one. If the TTC of the vehicle is less than the TTC threshold, which means a possible collision is recognised. The value of TTC can reflect the total number of such possible rear-end collisions for a given time period. Since the value of TTC is inversely related to crash risk, a expand index based on TTC is designed as a safety reward function in this work and it can prompt the DVSL agent to implement speed limit action to  reduce the collision risk of the traffic flow. To do so, the $NPC$ is defined as the number of vehicles with a potential collision and the value in the node surround at the time step $t$ (see  Eq. \ref{equ:ttc2}).

\begin{equation}
\begin{aligned}
NPC = & \sum_{i=0}^{M}TTC_{i} 
\\ & \text{when} \qquad  TTC_{i} < TTC_{c}
\end{aligned}
\label{equ:ttc2}
\end{equation}

\begin{equation}
\overline{NPC}= \frac{1}{T_c} \sum_{t=1}^{T_c} \frac{M-NPC(t)}{M}
\end{equation}

\noindent where $TTC_c$ is the threshold of the $TTC$, $M$ is the total number of vehicles. 

The reward function is defined as:
\begin{equation}
r = \frac{1}{2}(\overline{v}_{MA} + \overline{NPC})
\label{equ.reward}
\end{equation}

\subsubsection{\textbf{State}}

The DVSL controller in this paper provides the speed limits in each lane, so the occupancy and average speed of each lane will be considered in state space.
The \textbf{MA}  is the first to be affected by traffic bottlenecks, and the traffic capacity of the MA area will decrease;
If the implement of traffic control strategy is not in time, a large number of vehicles will accumulate and spread in upstream of the bottleneck point. Therefore, the state of the DVSL controller should focus on 5 areas: 1 is \textbf{MA} and  4 are \textbf{MI}, \textbf{DSA}, \textbf{AA} and \textbf{RI}, which is in the upstream of merging point. The state $\textbf{s}$ is denoted as Eq. \ref{equ:state}.

\begin{equation}
\textbf{s} = [s_{MI,1},s_{MI,2},\dots,s_{MA,N_{MA}} ]_{N_s}
\label{equ:state}
\end{equation}

\begin{equation}
N_s=N_{MI}+N_{DSA}+N_{AA}+N_{RI}+N_{MA}
\end{equation}

\begin{equation}
s_{MI,1} =\{(o_{MI}^{1},v_{MI}^{1})\}
\label{equ.occupation_speed}
\end{equation}

\noindent take the MI parameters as example, $s_{MI,1}$ represent the state of lane $1$ in MI; $N_s$ represent the number of total lanes in 5 area above mentioned; $N_{MI}$ is the number of lanes of MI; $o_{MI}^{1}$ is the occupation of lane $l$ of the MI; $v_{MI}^{1}$ is the average speed of lane $l$ of the MI;

In this work, the geometric features of the target traffic network are expressed by a graph and the traffic state data is embedded into the graph to represent the state of the proposed controller. The lanes in \textbf{MA}, \textbf{MI}, \textbf{DSA}, \textbf{AA} and \textbf{RI} are defined as graph nodes and the relationship between lane and lane as graph edges.

(1)~Graph

The state can be encoded into a directed graph, $\mathcal{G}_t^{(0)}=(V_t^{(0)}, E)$, where $V_t^{(0)}$ is a set of the initial node features for the graph, and $E$ is the set of graph edges which is defined as the relationship between lanes in this paper. The initial node features are initialized as follows:

\begin{equation}
V^{(0)}(t) = \{ V_1^{(0)}(t),V_2^{(0)}(t),\dots, V_i^{(0)}(t)\}
\end{equation}

\noindent where $i \in I$ is the number of nodes, which is equal to $N_s$. $v_i^{(0)}(t)$ is a traffic state representation of $I_th$ node and is the state of each lane, as is shown in Eq. \ref{equ.occupation_speed}. The value of occupancy and average speed in each lane indicates the node features of the input graph.

\begin{equation}
E = \{e_{i,j}\}_{i \neq j}
\label{equ.adjacency_matrix}
\end{equation}

\noindent where $E$ the adjacency matrix representing the graph structure; $e_{i,j}$ represents a relationship between two lanes $i$ and $j$ determined depending on the traffic flow direction. We classify the types of relationship between lanes into two groups:

\begin{itemize}
    \item $e_{ij} = 1$ if $i$ is upstream of $j$, or $i$, $j$ are neighbours.
    \item $e_{ij} = 0$ not match above condition.
\end{itemize}  

The relationships between lanes are determined, so once the adjacency matrix $E$ is established, it does not change with the passage of time.


(2)~Graph state representation

The initial node features for the graph are processed into an information-condensed node and relationship features of the target graph by employing graph message passing  \cite{yoon2021transferable}. 
A single update is composed of the following iterative computational steps.

\textbf{Step 1}: Message passing

In the message passing, the feature update of each lane is determined by collecting and aggregating information from its neighbouring lanes. This process can be expressed using the following formula:

\begin{equation}       
H= V \times E  \times W^{T} 
\label{equ.message_passing}
\end{equation}

\noindent where $H$ is the updated feature matrix; $E$ is the adjacency matrix representing the graph structure,as shown in Eq.\ref{equ.adjacency_matrix}. $V$ is the feature matrix containing the features of all nodes, as shown in  Eq.\ref{equ.occupation_speed}. $W$ is the weight matrix used for message passing. This formula captures the propagation of information from neighboring lanes to each lane in the graph structure of road network.

\textbf{Step 2}: Aggregation:

In the aggregation process, the updated feature representations from neighboring nodes are aggregated using the sigmoid function to generate a new feature representation for each node. 

\begin{equation}       
H^{'}=  sigmoid(H)
\label{equ.agreation}
\end{equation}

\noindent where the sigmoid function which squashes the input value to a range between 0 and 1. The sigmoid function introduces non-linearity and can help capture complex relationships between the aggregated features.

\subsection{Deep Reinforcement Learning Algorithm}

As mentioned previously, the actor-critic approach has been chosen for this work, and specifically, we have implemented the Proximal Policy Optimization (PPO) algorithm \cite{schulman2017proximal}, which employs multiple epochs of stochastic gradient ascent for each policy update. These methods are known for their stability and reliability.

Algorithm \ref{alg:DVSL_PPO} presents the training process of the DVSL controller.
The policy $\pi$ is designed to direct the DVSL controller in selecting the most suitable action for a given state $s$, with the goal of maximizing the cumulative reward value $r$.
We predefined the termination condition as the number of training iterations $I$. In each iteration, there are $E$ episodes running in parallel, and each episode lasts $H$ timesteps.
The trajectory data $\tau$ including $V_t,a_t,r_t$ is collected in $N$ timesteps, and the value of estimate advantages $\hat{A}$ is calculated based on the collecting trajectory $\tau$ (see line 10, Algorithm \ref{alg:DVSL_PPO}); 
update the policy parameter $\theta$ through gradient ascent with sampled $M$ timesteps  (see line 13, Algorithm \ref{alg:DVSL_PPO}) and then update the parameter $\omega$ through gradient descent with $B$ timesteps  (see line 17, Algorithm \ref{alg:DVSL_PPO});
To avoid instability in the process of policy update we have included the following measures:  if the current policy is better than the previous strategy, the gradient length will extend and if the current policy is not better than the previous strategy, the gradient length is less (see lines 20 to 23, Algorithm \ref{alg:DVSL_PPO}).

\begin{algorithm}
    \caption{Training process of DVSL using PPO}
    \label{alg:DVSL_PPO}
    \begin{algorithmic}[1]
        \REQUIRE
        \STATE Obtain the ID of DVSL controller.
        \STATE Set the number of episodes in parallel to $E$, and the time horizon for each episode to $N$
        \STATE Initialize the policy parameter for each agent, $\pi_\theta$; Initialize the value network parameter for each agent, $\omega$
        \STATE Initialize sample batch $B$, $M$ 
        \ENSURE
        \FOR {Iteration = 1,2,3,...,$I$}
            \FOR {episode = 1,2,3,...,$E$}
                \FOR {timesteps $t$ = 1,2,3,...,$N$}
                \STATE collecting the node feture $V$ and graph topology $E$ and than based on the Eq. \ref{equ.message_passing},\ref{equ.adjacency_matrix} in turn, the state $V_t$ with graph information can be obtained.
                \STATE Run policy $\pi_\theta$ for $T$ timesteps, collecting $\tau = \{V_t,a_t,r_t\}$
                \STATE Estimate advantages $\hat{A} = \sum_{t^{'}>t}\gamma^{t^{'}-t}r_{t^{t^{'}}}-V_{\omega}(V_{t})$
                \STATE $\pi_{old}\gets\pi_\theta$
                \FOR{ $j\in \{1,...,M\}$}
                \STATE $J_{PPO}(\theta)=\sum_{t=1}^{T}\frac{\pi_\theta(a_t|V_t)}{\pi_{old}(a_t|V_t)}\hat{A}-\lambda KL[\pi_{old}|\pi_\theta]$
                \STATE Update $\theta$ by a gradient ascent method w.r.t. $J_{PPO}(\theta)$ 
                \ENDFOR
                \FOR{$j\in \{1,...,B\}$}
                \STATE $L_{BL}(\omega) = -\sum_{t=1}^{T}\hat{A}^{2}$
                \STATE Update $\omega$ by a gradient decent method w.r.t.$L_{BL}(\omega)$
                \ENDFOR                   
                \IF{$KL[\pi_{old}|\pi_\theta]>\beta_{high}KL_{target}$}
                \STATE $\lambda\gets\alpha\lambda$
                \ELSIF{$KL[\pi_{old}|\pi_\theta]< \beta_{high}KL_{target}$}
                \STATE $\lambda\gets\frac{\lambda}{\alpha}$
                \ENDIF
                \ENDFOR
            \ENDFOR
        \ENDFOR
    \end{algorithmic}
\end{algorithm}

\section{Evaluation Methodology}\label{s3}

To assess the performance of the proposed algorithm, experiments are conducted on a high-performance computer with the following hardware configuration: Processor: AMD Ryzen 7 5800 8-Core (16 CPUs), 3.4GHz; Memory: 16384MB RAM; Graphics Card: NVIDIA GeForce RTX 3070. Python version 3.8 is used as the programming language. The simulation platform used is Simulation of Urban MObility (SUMO) version 1.16.0\footnote{\url{https://eclipse.dev/sumo/}}, which is one of the most widely used open-source microscopic traffic simulators. Our model design and implementation are based on SUMO-RL\footnote{\url{https://lucasalegre.github.io/sumo-rl/}}, which provides DRL-related API to work with SUMO dynamically. The algorithm part is from Ray[rllib] version 2.3.1\footnote{\url{https://docs.ray.io/en/latest/rllib/index.html}}, is an open-source library, offering support for production-level, highly distributed RL workloads.

\subsection{Testing Scenario}

\begin{figure}[!ht]
\centerline{\includegraphics[width=0.95\columnwidth]{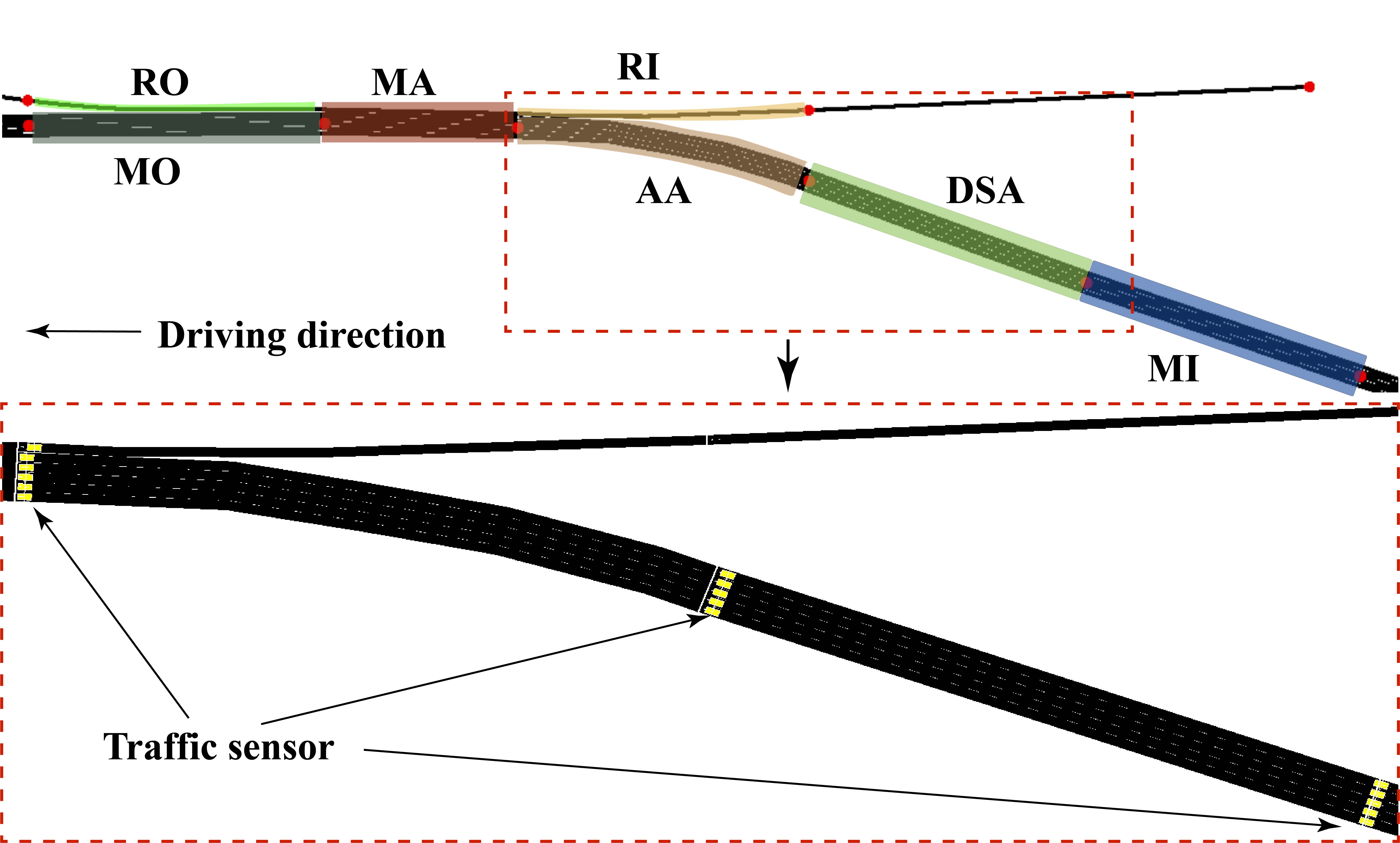}}
	\caption{Area divide and detectors deploy of state space}\label{fig.sce}
\end{figure}

The selection of the scenario is based on the study conducted in the literature\cite{wu2020differential}. It is a classic construct and the traffic network can be any freeway section with on and off-ramps. 
In the V2I environment, $\Delta t$ is updated every 5 seconds, and the free flow speed is 100 km/h
Considering that the segment length should satisfy $\Delta x \geq v_{seg}\Delta t$ according to the CFL condition \cite{wang2022integrated}, 200 meters is set as the segment length. 
In addition, vehicle detection devices are deployed on each lane at every 200 meters upstream of the bottleneck area, which is the main source nformation of state space. The induction loop detectors ($E1$) are used and collect information every 30-time steps. The details of the area divide and the deployment of detection devices are shown in Fig. \ref{fig.sce}. In this scenario, there are 5 lanes in \textbf{MI}, \textbf{DSA} and \textbf{AA}, respectively; there is 1 lane in \textbf{RI} area; there are 6 lanes in \textbf{MA} area. Each lane has two types of traffic information: occupancy and mean speed. Therefore, 44 states and 5 actions in this paper.

\begin{figure}[!ht]
    \centerline{\includegraphics[width=0.99\linewidth]{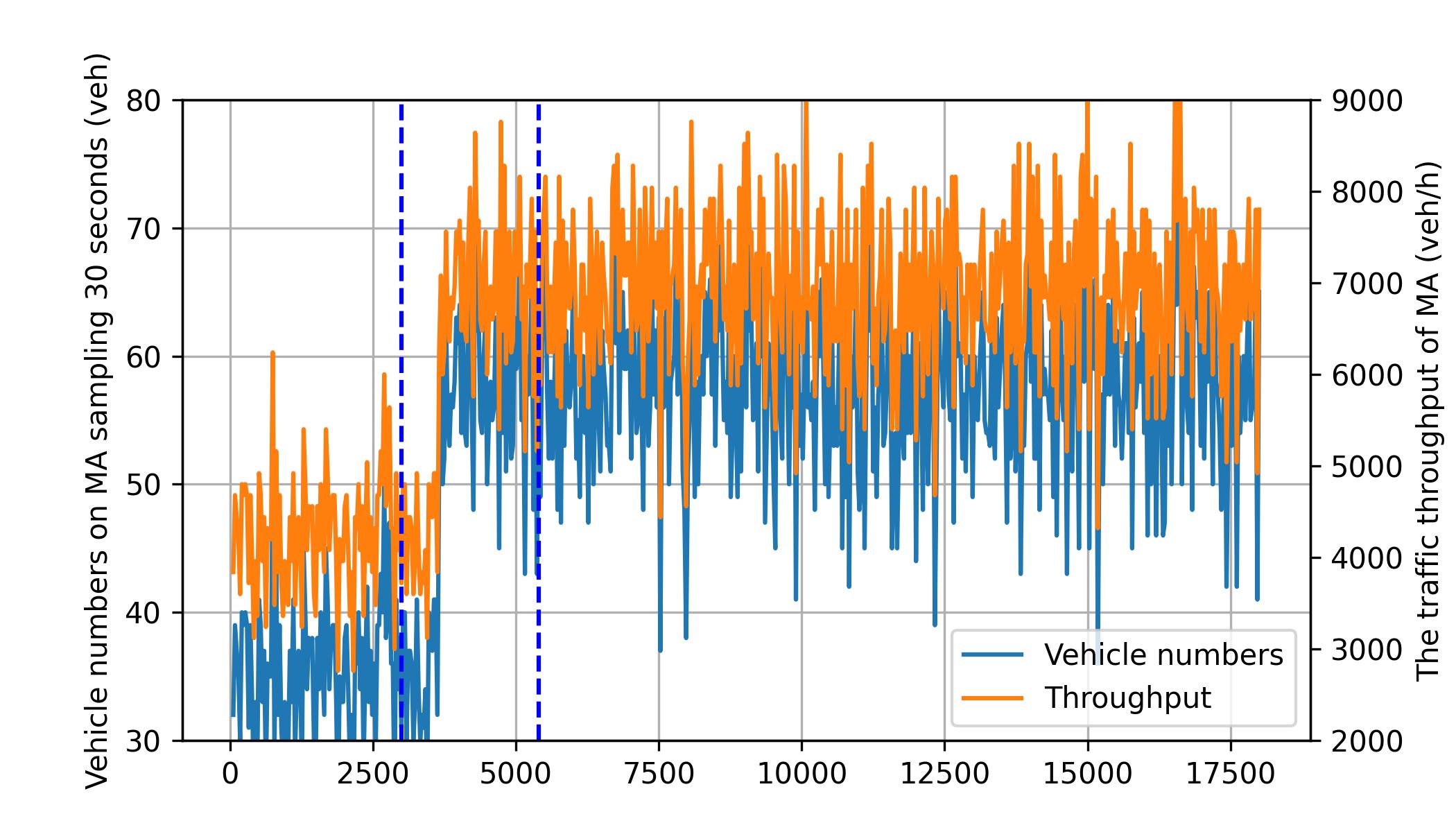}}
	\caption{The traffic traffic performance of MA.}\label{fig.no_bt}
\end{figure}

The traffic generated is also set to be consistent with the literature \cite{wu2020differential} and includes three route choices: (1) From mainlane to mainlane (M2M), (2) From mainlane to off-ramp (M2Off), and (3) From on-ramp to mainlane (On2M). The vehicle depart will last for 5 hours from 5:00 am to 10:00 am. The number of vehicles per hour, within three routes, is modelled as a Poisson process. Similar to the literature, each simulation lasts 18000 seconds \cite{wu2020differential}. The number of vehicles is collected by the traffic detectors sampling every 30 seconds, as shown by the blue line in Fig.\ref{fig.no_bt}. Based on the sampling data, the throughput can be calculated and is shown by the orange line in Fig.\ref{fig.no_bt}. 
As shown in  Fig.\ref{fig.no_bt}, there is an obvious increase in throughput at \textbf{MA} at around 3200 seconds and then it fluctuates at around 7000 $veh/h$. Based on the observation, there is a traffic bottleneck after 3200 seconds. To focus on solving the bottleneck problem, the training process for DVSL will commence at 3000 seconds and conclude at 5400 seconds, lasting for a duration of 2400 seconds.

In order to make the traffic environment more complex, there are 4 types of vehicles in traffic flow, as is shown in Table \ref{tab:veh_info}. The $speedFactor$ is applied in each and is the mean and deviation of the speed distribution of traffic flow, which is beneficial to the realism of a simulation. The parameter $lcSpeedGain$ represents the extent to which vehicles adjust lanes to increase speed. In this paper,  vehicles may switch to an adjacent lane with a higher speed limit to achieve higher speeds when the speed limit of a lane is low. This prevents vehicles from being overly conservative and lingering in lanes with low-speed limits for extended periods.


\begin{table}
\centering
\caption{Types of vehicles\label{tab:veh_info}}
\begin{tabular}{p{1.8cm}|p{0.9cm}<{\centering}p{0.9cm}<{\centering}p{0.9cm}<{\centering}p{0.9cm}<{\centering}} 
\toprule 
 Paremeters  &Type 1    &Type 2      &Type 3       &Type 4        \\ 
\midrule 
Length  &8	&8 	&3.5 	&3.5 \\ 
\midrule 
carFollowModel  &Krauss &IDM 	&Krauss &IDM \\ 
\midrule 
speedFactor & \multicolumn{4}{c}{normc(1,0.1)} \\
\midrule 
lcSpeedGain  &1 &0.8 &1 &0.8 \\
\bottomrule 
\end{tabular}
\end{table}

\subsection{Evaluation Metrics}

\begin{itemize}
    \item \textbf{Efficiency}:
    Average waiting time(AWT) ($second$). The sum of all vehicles' consecutive standing  which means the speed below 0.1 $m/s$;
    Total Stop Times (TST) ($veh$) denotes the aggregate count of vehicles moving at speeds below 0.1 $m/s$.
    Bottleneck Throughput (BT) ($veh/h$) represents the number of vehicles passing the bottleneck area during one hour.
    \item \textbf{Safety}:  
    TTC which means a possible collision is recognized when the time gap between the two adjacent cars is less than the threshold of 3 seconds. The sum of the over threshold value is the $NPC$ which is calculated by Eq. \ref{equ:ttc2}.  
\end{itemize}

\subsection{Compared Methods}

To test the effectiveness of our proposed system, we compared the performance of the following algorithms.

\begin{itemize}
    \item \textbf{Baseline}: The baseline is NO-VSL control in the on-ramp area. The partial traffic performance is shown in \ref{fig.no_bt}, proving that without traffic management and control measures, the MA will quickly reach the maximum throughput, causing a traffic bottleneck.
    \item \textbf{DVS-Rule}: DVS-Rule (Rule-based DVSL) is similar to VS-Rule. The VSL signs automatically change according to traffic occupancy and the flow of each lane in the bottleneck area \cite{yuan2022selection}. 
    \item \textbf{DVS-DDPG}: DVS-DDPG method \cite{wu2020differential} is a state-of-the-art DRL-based, in which dynamic and distinct speed limits among lanes can be imposed. 
    The proposed DRL model uses DDPG based on a novel actor-critic architecture to learn a large number of discrete speed limits in a continuous action space. The action in Eq.\ref{equ:action}, state space in Eq.\ref{equ:state} and reward in Eq.\ref{equ:state} are used. 
    \item \textbf{TD3LVSL}: TD3LDVSL method \cite{lu2023td3lvsl} is another state-of-the-art DRL-based, which is a reinforcement learning-based lane-level VSL (LVSL) control approach for conducting refined traffic control on the mainlane. An actor-critic framework is developed to generate and evaluate the discrete speed limits of each lane in continuous action space. The action in Eq.\ref{equ:action}, state space in Eq.\ref{equ:state} and reward in Eq.\ref{equ:state} are used.   
    \item \textbf{DVS-TD3}: DVS-TD3 method is a TD3-based DVSL control. The state space is as shown in Eq.\ref{equ:state}.  DVS-TD3 aims to demonstrate that the PPO algorithm has a good performance on DVSL control compared to TD3.
    \item \textbf{DVS-PPO}: The DVS-PPO method is a DVSL control based on the PPO algorithm. The state space is as shown in Eq.\ref{equ:state}. Introducing DVS-PPO aims to demonstrate whether the state with topology has an advantage in DRL training process.
    \item \textbf{DVS-RG}: DVS-RG (DVSL based on DRL with Graph State Representation) is proposed in this paper. DVS-RG and DVS-PPO are based on the PPO algorithm. The two methods have the same reward and action. The difference is that the state space of the DVS-PPO method is a matrix represented by occupancy and speed, as shown in Eq.\ref{equ:state}, but the occupancy and speed in a directed graph are integrated as the state space, as shown in Eq. \ref{equ.message_passing}, \ref{equ.agreation}.
\end{itemize}

All the above-mentioned DRL methods have the same action, reward and state, except for TD3LVSL. 
The reward training process of the controllers of VS-PPO, DVS-PPO, DVS-TD3, DVS-DDPG and DVS-RG methods are shown in Fig. \ref{fig.training}. The design of reward and in TD3LVSL is based on the work presented in \cite{lu2023td3lvsl} and the training process of all is shown in the bottom part of Fig. \ref{fig.training}.  

The algorithms of VS-PPO, DVS-PPO, DVS-TD3 and DVS-RG are from RLLIB which is open-resource. The algorithm of DVS-DDPG is from \cite{wu2020differential} which is a customised actor-critic architecture. 

\section{Evaluation Results and Analysis}\label{s4}

\subsubsection{Comparison With Other RL Methods}

In this section, 5 DRL-based methods are tested in the scenario depicted in  Fig. \ref{fig.sce}. 

\begin{figure}[!ht]
	\centerline{\includegraphics[width=0.99\columnwidth]{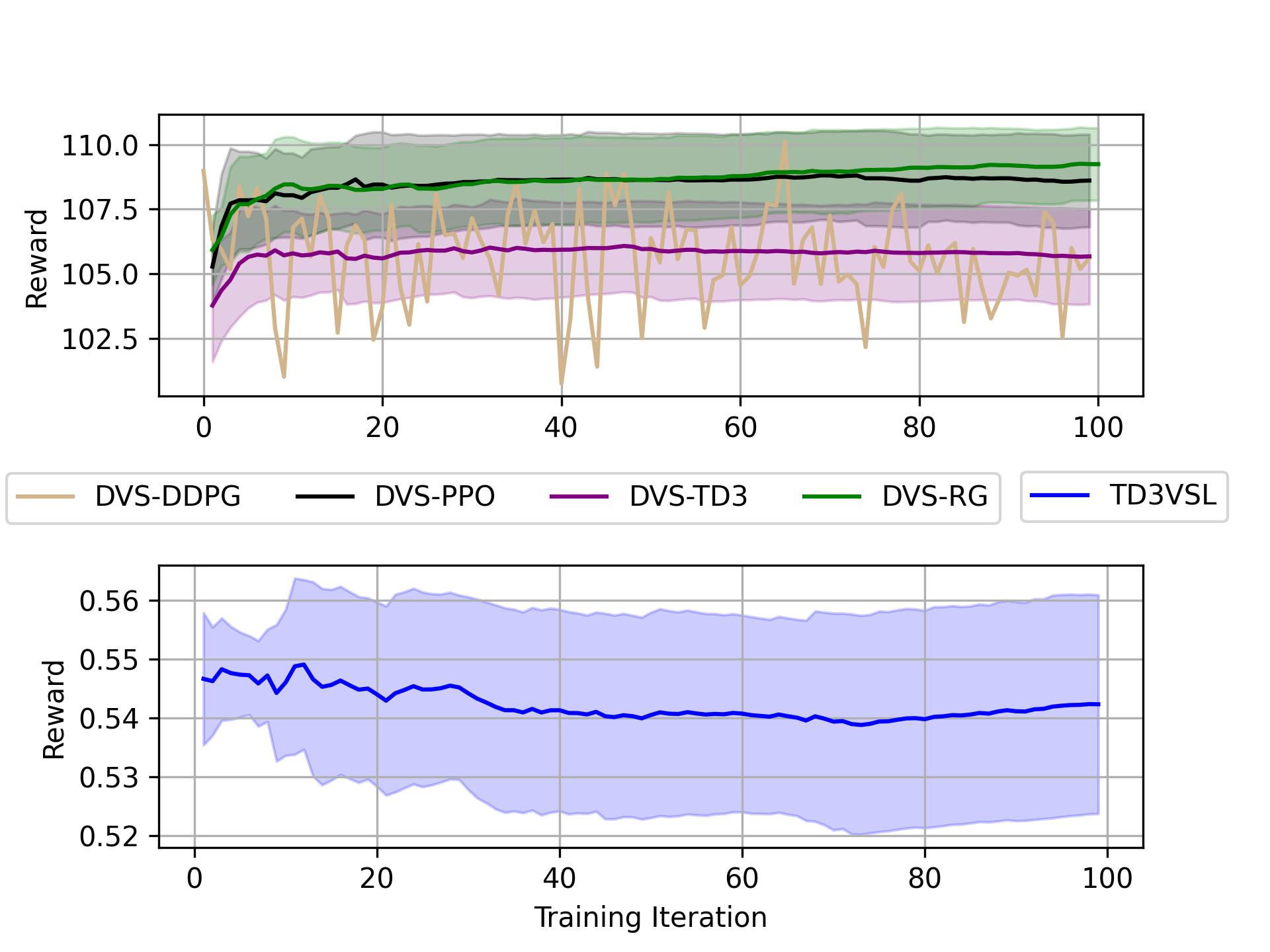}}
	\caption{Training process of DVS-PPO, DVS-TD3, DVS-DDPG and DVS-RG methods.}\label{fig.training}
\end{figure}

From Fig.\ref{fig.training}, we can see the fluctuation of DVS-DDPG under the training process. The curve of reward has the same trend in \cite{wu2020differential} and is more difficult to converge. The reason predicted is that the customised model is more sensitive to hyperparameters. Compared with the DVS-DDPG, the reward of The VS-PPO, DVS-PPO, DVS-TD3 and DVS-RG grows steadily during the training process and are trained in the proposed environment framework. The The policy capacity is 100 in the default setting in this framework which means can be shared among the policies. In addition, there are 2 rollout workers during training, process, so the trajectory collection is double. That is the reason why is the smoothness of the training curve of VS-PPO, DVS-PPO, DVS-TD3 and DVS-RG. 

In Fig.\ref{fig.training}, the solid line indicates the mean reward value of each episode, and the shadowed area indicates the standard deviation.


We further compared the cumulative rewards of DRL-based VSL in Fig. \ref{fig.training}. Our goal is to maximize the reward in Eq. \ref{equ.reward}, and the higher reward implies a higher travelling speed and fewer potential collisions. We can see that both the PPO algorithms have the highest cumulative return during the training process, and the reward value curve has an upward reward trend before 40 episodes. The DVS-RG method has slightly outperformed the DVS-PPO method after 60 episodes and the reward curve of the training process increased slightly. The only difference between the two methods is that the state space of the DVS-PPO is a matrix with traffic status, while the state space of the DVS-RG contains not only the state information of the nodes but also the state of the leading node. This result shows that having the state information of the neighbour node is beneficial for the training of the DVSL control strategy.

Additionally, we observe fluctuations in the training process of TD3LVSL depicted in the lower section of Fig. \ref{fig.training}. We attribute this to two main factors: firstly, TD3LVSL\cite{lu2023td3lvsl} employs a complex action space consists 15, organized into 3 segments with 3 lanes each; secondly, the state space solely focuses on density. We believe these factors hinder the agent's ability to optimize the policy.


\begin{table}[]
\caption{Average performance under different methods.
\label{tab:result1}}
\begin{center}
\begin{threeparttable}   
\begin{tabular}{p{1.5cm}|p{1.2cm}<{\centering}p{1.2cm}<{\centering}p{1.2cm}<{\centering}|p{1.2cm}<{\centering}} 
\toprule 
Method & TST      & AWT      & BT      & TTC          \\
\midrule 
No-VSL                  & 518.00   & 8.86     & 6133.50 & 4574.00     \\
\midrule 
\multirow{2}{*}{DVS-Rule}  & 400.00   & 4.16     & 6028.50 & 4377.00     \\
                           & -22.78\% & -53.06\% & 0.05\%  & -4.31\%  \\
\midrule 
\multirow{2}{*}{DVS-DDPG} & 375.00   & 4.92     &6022.70 & 4159.25     \\
                           & -27.61\% & -44.43\% & 0.39\%  & -9.07\%   \\
\midrule 
\multirow{2}{*}{TD3LVSL}  & 378.38   & 4.19     & 6160.13 & 4435.25     \\
                           & -26.95\% & -52.72\% & 0.43\%  & -3.03\%   \\
\midrule 
\multirow{2}{*}{DVS-TD3}  & 332.75   & 3.30     & 6161.07 & 4285.50     \\
                           & -35.76\% & -62.76\% & 0.45\%  & -6.31\%   \\
\midrule 
\multirow{2}{*}{DVS-PPO}  &399.50 	&5.44 	&6151.01 	&3959.66 	 \\
                            &-22.88\%	&-38.56\%	&0.29\%	&-13.43\%	 \\
\midrule 
\multirow{2}{*}{DVS-RG} & \textbf{308.67} 	& \textbf{2.80} 	& \textbf{6175.76} 	& \textbf{3845.33} 	    \\
                            & \textbf{-40.41\%}	&\textbf{-68.44\%}	& \textbf{0.69\%} & \textbf{-15.93\%}   \\                     
\bottomrule 
\end{tabular}
\begin{tablenotes}    
    \footnotesize               
    \item The best performances are in bold	
    \end{tablenotes}            
    \end{threeparttable}       
    \end{center}
\end{table}

\begin{figure} 
  \centering
  \subfloat[\label{fig.TST}]{%
  \includegraphics[width=0.45\linewidth]{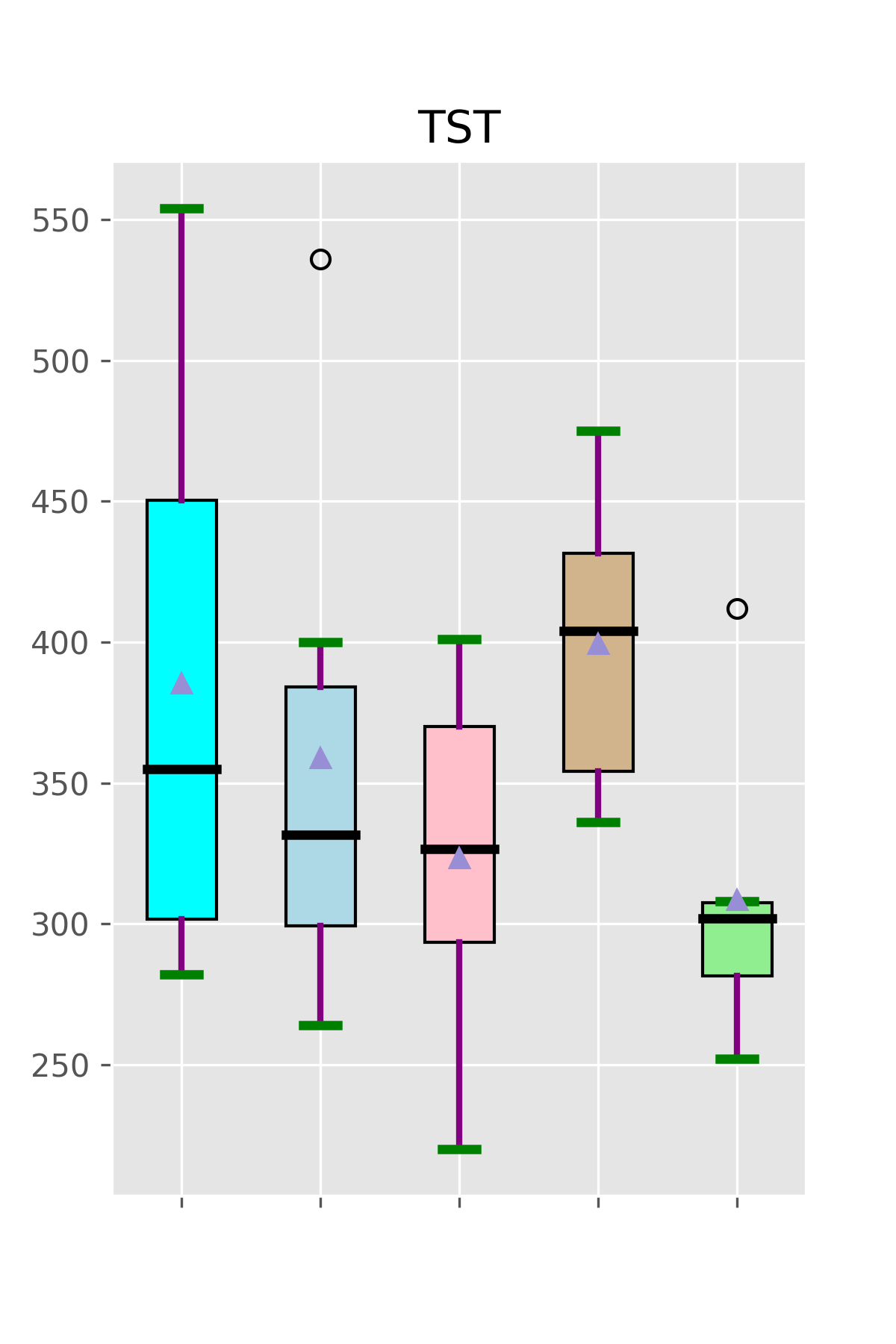}}
  \hfill
  \subfloat[\label{fig.AWT}]{%
  \includegraphics[width=0.45\linewidth]{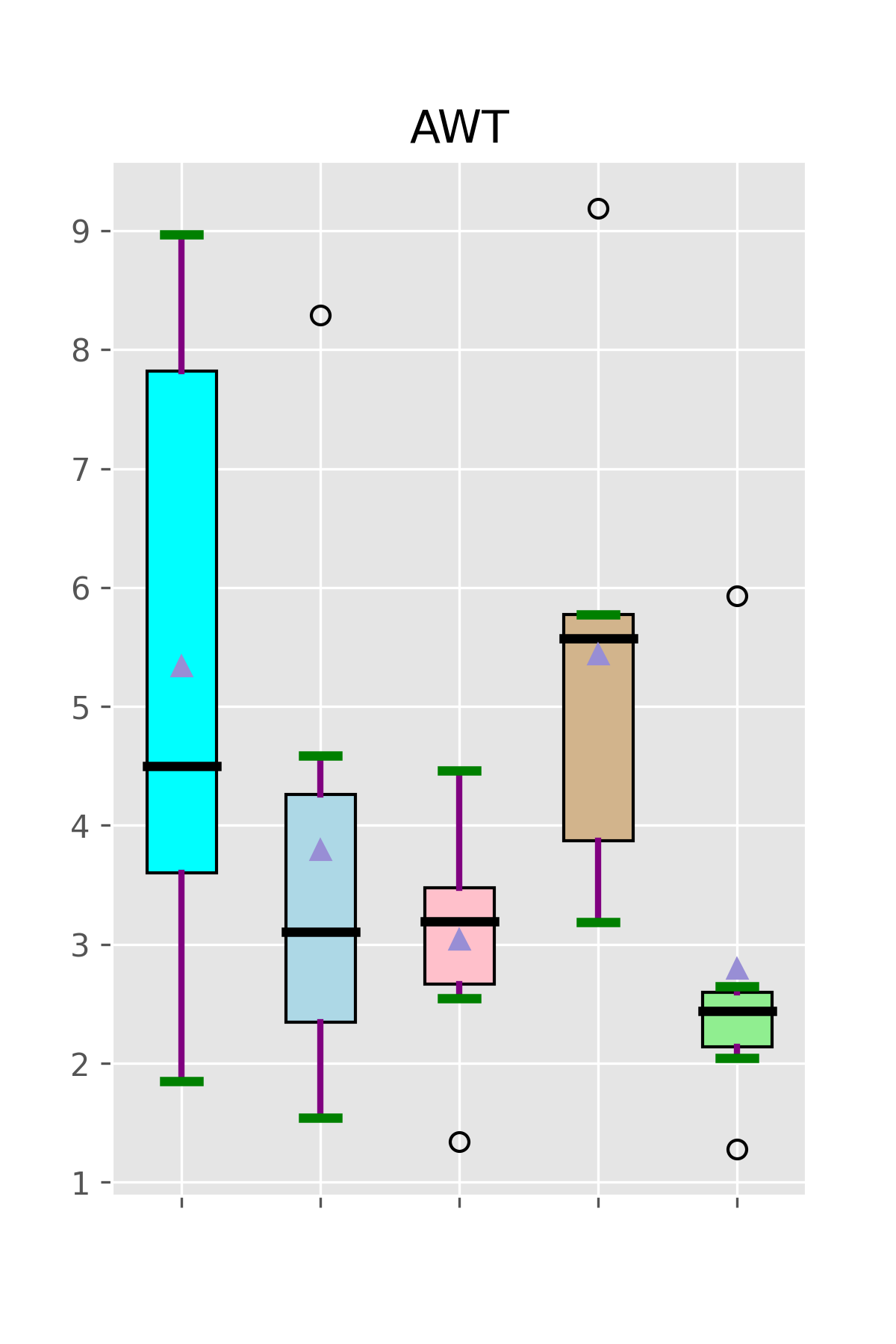}}
    \\
  \subfloat[\label{fig.BT}]{%
  \includegraphics[width=0.45\linewidth]{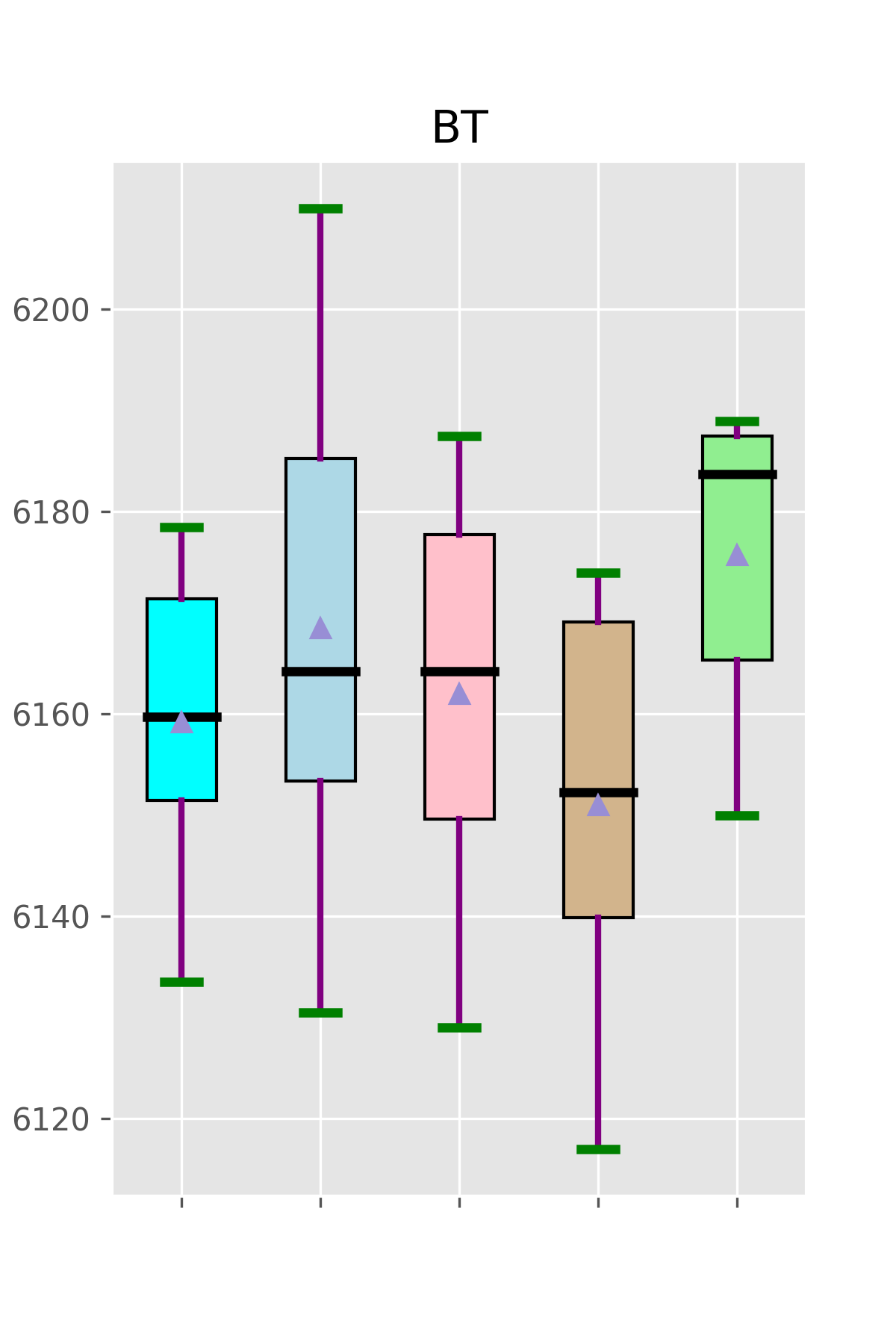}}
  \hfill
  \subfloat[\label{fig.TTC}]{%
  \includegraphics[width=0.45\linewidth]{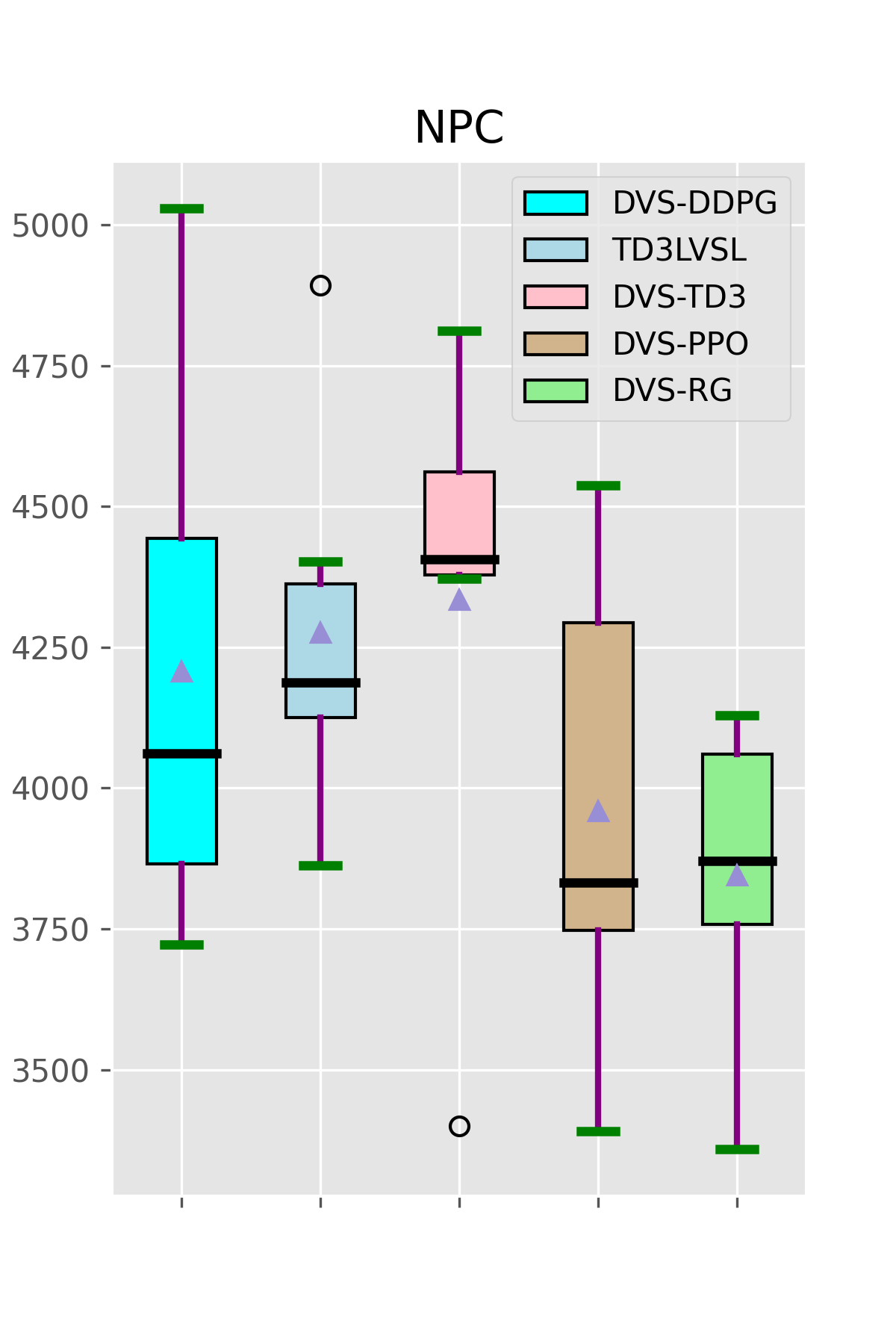}}
  \caption{The evaluation matrix distributions of 8 episodes;(a) TST;(b) AWT; (c) BT; (d) NFC}
  \label{fig1} 
\end{figure}

\subsubsection{Comparison With State-of-the-Art Methods}

Table \ref{tab:result1} presents the traffic performance across various methods within the same scenario, focusing on both efficiency and safety. The baseline is No-DVSL, where no traffic control method is implemented under a traffic bottleneck environment.
Figures \ref{fig.TST}, \ref{fig.AWT}, \ref{fig.BT}, and \ref{fig.TTC} show the distribution of the evaluation index across 8 episodes for each method, encompassing 4 state-of-the-art DRL-based algorithms along with our proposed method.

\textbf{Efficiency}: As shown in Table \ref{tab:result1}, in terms of TST, the DVS-RG method has an average reduction of 40.41\% compared with NO-VSL and has a more densely distributed TST values, as shown in Fig. \ref{fig.TST}; 
in terms of AWT, the DVS-RG method showed the most significant improvement at 68.44\%. The average AWT value of DVS-TD3 is close to the result of DVS-RG and is the second-best performance. 
From the Fig. \ref{fig.AWT}, the distribution of DVS-TD3 is the most densely expected DVS-RG. 
In terms of BT, all methods have some limited improvement.  The reasons for seeing these results could be related to the settings for the generated fixed traffic flow. Despite all this, DVS-RG has the best performance, increasing 0.69\% throughput in the merging area during the control process under the simulations compared to the baseline. Fig. \ref{fig.BT} shows that the dense distribution of BT for all DRL methods follows a similar distribution. Overall, the DRL-based VSL controllers have an advantage over the traditional method DVS-rule. DVS-RG performs best in terms of efficiency, followed by DVS-TD3. The efficiency results show that DVS-RG is better than DVS-PPO and we can conclude that state space representation is beneficial for agent's learning. Fig.\ref{fig.TST}, Fig.\ref{fig.AWT} and Fig.\ref{fig.BT} show the value distribution of results obtained by TD3LVSL, DVS-TD3, DVS-PPO and DVS-RG is stable compared with DVS-DDPG. In comparison to the DRL network we utilize, the DDPG with the customized neutral network has poor robustness.

\textbf{Safety}: DVS-RG method reduced TTC by average over 15.93\% of average each episode, as shown in Table \ref{tab:result1}. Other methods expect DVS-PPO to improve less traffic safety as well, which demonstrates that PPO-based traffic control measures are beneficial for the safety of road networks. This point can also be confirmed in Fig.\ref{fig.TTC}.

In summary, DVS-RG demonstrates superior performance in terms of efficiency and safety compared with no matter traditional methods and state-of-the-art methods. Expect DVS-DDPG, the other methods have a gentle reward curve and a dense distribution of results which demonstrates the proposed environment framework has the advantage on stable learning during the training process. The comparison of various indicators between TD3DVSL and TD3VSL can also prove the superiority of the proposed DRL architecture including the design of reward and state space. DVS-TD3 and DVS-PPO, with the same environmental framework, have differences that result in efficiency and safety. From the result shown TD3 algorithm is sensitive to efficiency reward and the PPO algorithm is sensitive to safety reward. This result can give some suggestions for researchers and engineers for strategy determination. DVS-RG outperforms DVS-PPO, which demonstrates the state space integrates the road network topology structure and traffic flow information have benefits for the agent to rapidly learn the relationship between the features resulting from the spatial structure. 


\section{Conclusions and Future Work}\label{s5}

This paper proposed a novel DVSL control strategy with DRL that was supported by a graph-based state representation, DVS-RG, to improve freeway traffic mobility and alleviate recurring bottlenecks by customising speed limits at proper locations based on the traffic state. DVS-RG has shown significant improvements in traffic efficiency (i.e., average waiting time, total stopped time and bottleneck throughput) as well as traffic safety (i.e., time-to-time collision ), which outperforms other DRL-based and traditional methods. Compared to the No-VSL method, DVS-RG can save up to 68.44\% in average waiting time while reducing potential collision time by up to 15.93\%.

In the future, experiments for larger transportation networks will be conducted to analyze the performance. Furthermore, the present work assumes data availability but some events (e.g. sensor failures, emergency traffic) could generate various types of anomalous data. Evaluating the robustness of the proposed method under several anomalous data may constitute the object of future studies.


\section*{Acknowledgment}

The authors would like to thank the Project (202206567025) supported by the China Scholarship Council.


\ifCLASSOPTIONcaptionsoff
  \newpage
\fi

\bibliographystyle{IEEEtran}
\bibliography{vsl_drl_reference}

\begin{thebibliography}{10}
\providecommand{\url}[1]{#1}
\csname url@samestyle\endcsname
\providecommand{\newblock}{\relax}
\providecommand{\bibinfo}[2]{#2}
\providecommand{\BIBentrySTDinterwordspacing}{\spaceskip=0pt\relax}
\providecommand{\BIBentryALTinterwordstretchfactor}{4}
\providecommand{\BIBentryALTinterwordspacing}{\spaceskip=\fontdimen2\font plus
\BIBentryALTinterwordstretchfactor\fontdimen3\font minus
  \fontdimen4\font\relax}
\providecommand{\BIBforeignlanguage}[2]{{%
\expandafter\ifx\csname l@#1\endcsname\relax
\typeout{** WARNING: IEEEtran.bst: No hyphenation pattern has been}%
\typeout{** loaded for the language `#1'. Using the pattern for}%
\typeout{** the default language instead.}%
\else
\language=\csname l@#1\endcsname
\fi
#2}}
\providecommand{\BIBdecl}{\relax}
\BIBdecl

\bibitem{chen2014variable}
D.~Chen, S.~Ahn, and A.~Hegyi, ``Variable speed limit control for steady and
  oscillatory queues at fixed freeway bottlenecks,'' \emph{Transportation
  Research Part B: Methodological}, vol.~70, pp. 340--358, 2014.

\bibitem{wang2022integrated}
C.~Wang, Y.~Xu, J.~Zhang, and B.~Ran, ``Integrated {{Traffic Control}} for
  {{Freeway Recurrent Bottleneck Based}} on {{Deep Reinforcement Learning}},''
  \emph{IEEE Transactions on Intelligent Transportation Systems}, vol.~23,
  no.~9, pp. 15\,522--15\,535, 2022.

\bibitem{wu2020differential}
Y.~Wu, H.~Tan, L.~Qin, and B.~Ran, ``Differential variable speed limits control
  for freeway recurrent bottlenecks via deep actor-critic algorithm,''
  \emph{Transportation Research Part C: Emerging Technologies}, vol. 117, p.
  102649, 2020.

\bibitem{wang2021freeway}
Y.~Wang, X.~Yu, S.~Zhang, P.~Zheng, J.~Guo, L.~Zhang, S.~Hu, S.~Cheng, and
  H.~Wei, ``Freeway {{Traffic Control}} in {{Presence}} of {{Capacity Drop}},''
  \emph{IEEE Transactions on Intelligent Transportation Systems}, vol.~22,
  no.~3, pp. 1497--1516, 2021.

\bibitem{lu2023td3lvsl}
W.~Lu, Z.~Yi, Y.~Gu, Y.~Rui, and B.~Ran, ``{{TD3LVSL}}: {{A}} lane-level
  variable speed limit approach based on twin delayed deep deterministic policy
  gradient in a connected automated vehicle environment,'' \emph{Transportation
  Research Part C: Emerging Technologies}, vol. 153, p. 104221, 2023.

\bibitem{yuan2022selection}
T.~Yuan, F.~Alasiri, and P.~A. Ioannou, ``Selection of the {{Speed Command
  Distance}} for {{Improved Performance}} of a {{Rule-Based VSL}} and {{Lane
  Change Control}},'' \emph{IEEE Transactions on Intelligent Transportation
  Systems}, vol.~23, no.~10, pp. 19\,348--19\,357, 2022.

\bibitem{khondaker2015variable}
B.~Khondaker and L.~Kattan, ``Variable speed limit: An overview,''
  \emph{Transportation Letters}, vol.~7, no.~5, pp. 264--278, 2015.

\bibitem{weikl2013traffic}
S.~Weikl, K.~Bogenberger, and R.~L. Bertini, ``Traffic management effects of
  variable speed limit system on a {{German Autobahn}}: {{Empirical}}
  assessment before and after system implementation,'' \emph{Transportation
  research record}, vol. 2380, no.~1, pp. 48--60, 2013.

\bibitem{jin2015control}
H.-Y. Jin and W.-L. Jin, ``Control of a lane-drop bottleneck through variable
  speed limits,'' \emph{Transportation Research Part C: Emerging Technologies},
  vol.~58, pp. 568--584, 2015.

\bibitem{du2019variable}
S.~Du and S.~Razavi, ``Variable {{Speed Limit}} for {{Freeway Work Zone}} with
  {{Capacity Drop Using Discrete-Time Sliding Mode Control}},'' \emph{Journal
  of Computing in Civil Engineering}, vol.~33, no.~2, p. 04019001, 2019.

\bibitem{karafyllis2019feedback}
I.~Karafyllis and M.~Papageorgiou, ``Feedback control of scalar conservation
  laws with application to density control in freeways by means of variable
  speed limits,'' \emph{Automatica}, vol. 105, pp. 228--236, 2019.

\bibitem{carlson2011local}
R.~C. Carlson, I.~Papamichail, and M.~Papageorgiou, ``Local {{Feedback-Based
  Mainstream Traffic Flow Control}} on {{Motorways Using Variable Speed
  Limits}},'' \emph{IEEE Transactions on Intelligent Transportation Systems},
  vol.~12, no.~4, pp. 1261--1276, 2011.

\bibitem{muller2015microsimulation}
E.~R. M{\"u}ller, R.~C. Carlson, W.~Kraus, and M.~Papageorgiou,
  ``Microsimulation {{Analysis}} of {{Practical Aspects}} of {{Traffic Control
  With Variable Speed Limits}},'' \emph{IEEE Transactions on Intelligent
  Transportation Systems}, vol.~16, no.~1, pp. 512--523, 2015.

\bibitem{mao2022variable}
P.~Mao, X.~Ji, X.~Qu, L.~Li, and B.~Ran, ``A {{Variable Speed Limit Control
  Based}} on {{Variable Cell Transmission Model}} in the {{Connecting Traffic
  Environment}},'' \emph{IEEE Transactions on Intelligent Transportation
  Systems}, vol.~23, no.~10, pp. 17\,632--17\,643, 2022.

\bibitem{guo2020integrated}
Y.~Guo, H.~Xu, Y.~Zhang, and D.~Yao, ``Integrated {{Variable Speed Limits}} and
  {{Lane-Changing Control}} for {{Freeway Lane-Drop Bottlenecks}},'' \emph{IEEE
  Access}, vol.~8, pp. 54\,710--54\,721, 2020.

\bibitem{han2021lineara}
Y.~Han, M.~Wang, Z.~He, Z.~Li, H.~Wang, and P.~Liu, ``A linear {{Lagrangian}}
  model predictive controller of macro- and micro- variable speed limits to
  eliminate freeway jam waves,'' \emph{Transportation Research Part C: Emerging
  Technologies}, vol. 128, p. 103121, 2021.

\bibitem{ye2020modelfree}
Y.~Ye, D.~Qiu, X.~Wu, G.~Strbac, and J.~Ward, ``Model-{{Free Real-Time
  Autonomous Control}} for a {{Residential Multi-Energy System Using Deep
  Reinforcement Learning}},'' \emph{IEEE Transactions on Smart Grid}, vol.~11,
  no.~4, pp. 3068--3082, 2020.

\bibitem{guo2023cotv}
J.~Guo, L.~Cheng, and S.~Wang, ``{{CoTV}}: {{Cooperative Control}} for
  {{Traffic Light Signals}} and {{Connected Autonomous Vehicles Using Deep
  Reinforcement Learning}},'' \emph{IEEE Transactions on Intelligent
  Transportation Systems}, pp. 1--12, 2023.

\bibitem{yang2021automaticb}
J.~Yang, P.~Wang, W.~Yuan, Y.~Ju, W.~Han, and J.~Zhao, ``Automatic generation
  of optimal road trajectory for the rescue vehicle in case of emergency on
  mountain freeway using reinforcement learning approach,'' \emph{IET
  Intelligent Transport Systems}, vol.~15, no.~9, pp. 1142--1152, 2021.

\bibitem{silver2016mastering}
D.~Silver, A.~Huang, C.~J. Maddison, A.~Guez, L.~Sifre, G.~{van den Driessche},
  J.~Schrittwieser, I.~Antonoglou, V.~Panneershelvam, M.~Lanctot, S.~Dieleman,
  D.~Grewe, J.~Nham, N.~Kalchbrenner, I.~Sutskever, T.~Lillicrap, M.~Leach,
  K.~Kavukcuoglu, T.~Graepel, and D.~Hassabis, ``Mastering the game of {{Go}}
  with deep neural networks and tree search,'' \emph{Nature}, vol. 529, no.
  7587, pp. 484--489, 2016.

\bibitem{bai2022hybrid}
Z.~Bai, P.~Hao, W.~ShangGuan, B.~Cai, and M.~J. Barth, ``Hybrid {{Reinforcement
  Learning-Based Eco-Driving Strategy}} for {{Connected}} and {{Automated
  Vehicles}} at {{Signalized Intersections}},'' \emph{IEEE Transactions on
  Intelligent Transportation Systems}, vol.~23, no.~9, pp. 15\,850--15\,863,
  2022.

\bibitem{kusic2020overview}
K.~Ku{\v s}i{\'c}, E.~Ivanjko, M.~Greguri{\'c}, and M.~Mileti{\'c}, ``An
  {{Overview}} of {{Reinforcement Learning Methods}} for {{Variable Speed Limit
  Control}},'' \emph{Applied Sciences}, vol.~10, no.~14, p. 4917, 2020.

\bibitem{walraven2016traffica}
E.~Walraven, M.~T.~J. Spaan, and B.~Bakker, ``Traffic flow optimization: {{A}}
  reinforcement learning approach,'' \emph{Engineering Applications of
  Artificial Intelligence}, vol.~52, pp. 203--212, 2016.

\bibitem{li2017reinforcement}
Z.~Li, P.~Liu, C.~Xu, H.~Duan, and W.~Wang, ``Reinforcement {{Learning-Based
  Variable Speed Limit Control Strategy}} to {{Reduce Traffic Congestion}} at
  {{Freeway Recurrent Bottlenecks}},'' \emph{IEEE Transactions on Intelligent
  Transportation Systems}, vol.~18, no.~11, pp. 3204--3217, 2017.

\bibitem{wang2019new}
C.~Wang, J.~Zhang, L.~Xu, L.~Li, and B.~Ran, ``A {{New Solution}} for {{Freeway
  Congestion}}: {{Cooperative Speed Limit Control Using Distributed
  Reinforcement Learning}},'' \emph{IEEE Access}, vol.~7, pp. 41\,947--41\,957,
  2019.

\bibitem{han2022new}
Y.~Han, A.~Hegyi, L.~Zhang, Z.~He, E.~Chung, and P.~Liu, ``A new reinforcement
  learning-based variable speed limit control approach to improve traffic
  efficiency against freeway jam waves,'' \emph{Transportation Research Part C:
  Emerging Technologies}, vol. 144, p. 103900, 2022.

\bibitem{kusic2021spatialtemporalc}
K.~Ku{\v s}i{\'c}, E.~Ivanjko, F.~Vrbani{\'c}, M.~Greguri{\'c}, and
  I.~Dusparic, ``Spatial-{{Temporal Traffic Flow Control}} on {{Motorways Using
  Distributed Multi-Agent Reinforcement Learning}},'' \emph{Mathematics},
  vol.~9, no.~23, p. 3081, 2021.

\bibitem{peng2022combined}
C.~Peng and C.~Xu, ``Combined variable speed limit and lane change guidance for
  secondary crash prevention using distributed deep reinforcement learning,''
  \emph{Journal of Transportation Safety \& Security}, vol.~14, no.~12, pp.
  2166--2191, 2022.

\bibitem{devailly2022igrla}
F.-X. Devailly, D.~Larocque, and L.~Charlin, ``{{IG-RL}}: {{Inductive Graph
  Reinforcement Learning}} for {{Massive-Scale Traffic Signal Control}},''
  \emph{IEEE Transactions on Intelligent Transportation Systems}, vol.~23,
  no.~7, pp. 7496--7507, 2022.

\bibitem{chen2021graph}
S.~Chen, J.~Dong, P.~Y.~J. Ha, Y.~Li, and S.~Labi, ``Graph neural network and
  reinforcement learning for multi-agent cooperative control of connected
  autonomous vehicles,'' \emph{Computer-Aided Civil and Infrastructure
  Engineering}, vol.~36, no.~7, pp. 838--857, 2021.

\bibitem{yoon2021transferable}
J.~Yoon, K.~Ahn, J.~Park, and H.~Yeo, ``Transferable traffic signal control:
  {{Reinforcement}} learning with graph centric state representation,''
  \emph{Transportation Research Part C: Emerging Technologies}, vol. 130, p.
  103321, 2021.

\bibitem{schulman2017proximal}
J.~Schulman, F.~Wolski, P.~Dhariwal, A.~Radford, and O.~Klimov, ``Proximal
  {{Policy Optimization Algorithms}},'' 2017.

\end{thebibliography}

\vspace{-5ex}

\begin{IEEEbiography}
[{\includegraphics[width=1in,height=1.25in,clip,keepaspectratio]{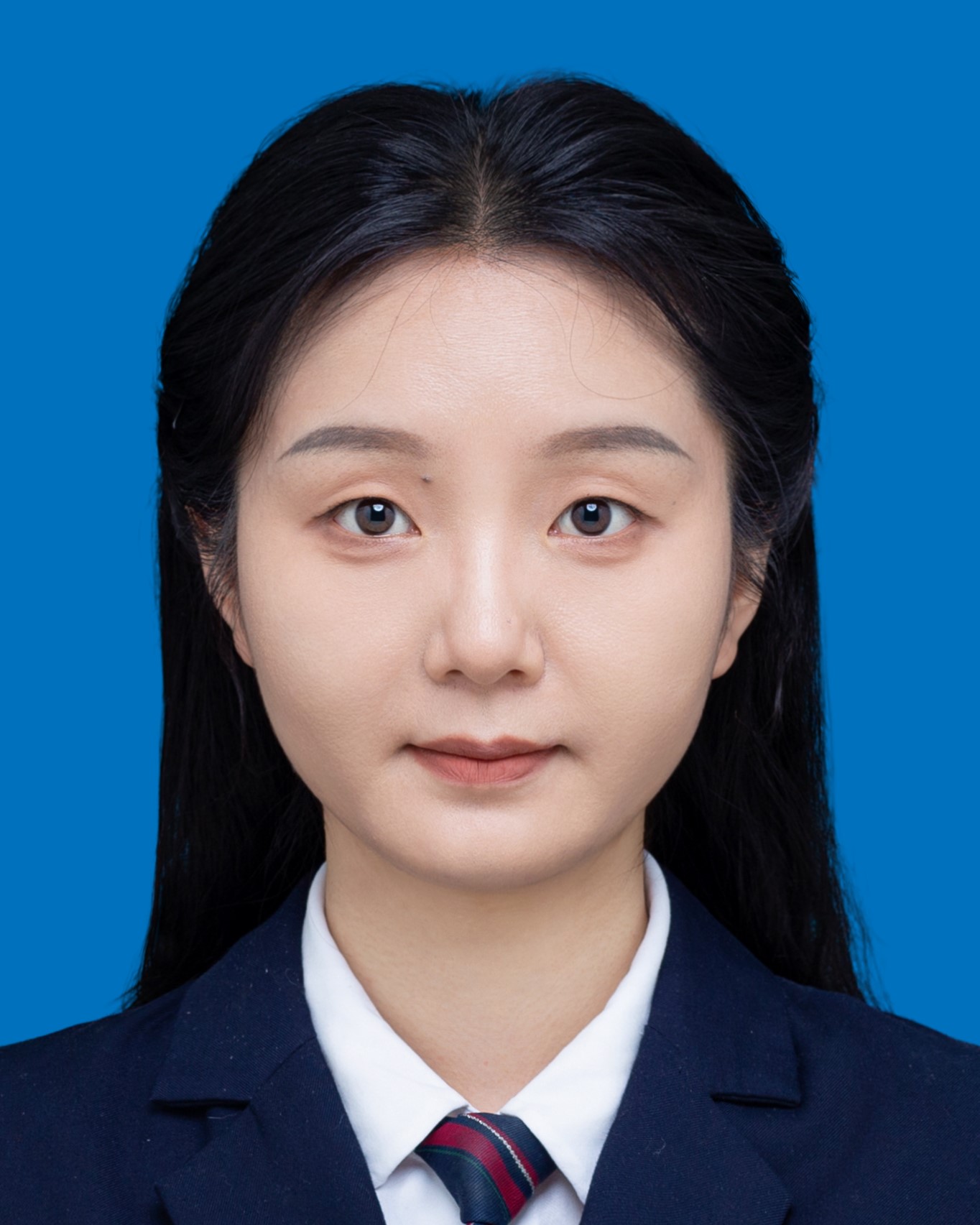}}]{Jingwen Yang} (Graduate Student Member, IEEE) received the B.S. degree from West Anhui University. Since 2018, She is pursuing a PhD degree in Transportation Information Engineering and Control at Chang'an University, China. She was a visiting student at the University College Dublin, Ireland, from 2022 to 2023. She was a recipient of the IEEE Intelligent Transportation Systems Society Young Professionals Travelling Fellowship in 2023. Her current research interests include intelligent transportation systems, and reinforcement learning and its applications.
\end{IEEEbiography}

\vspace{-5ex}

\begin{IEEEbiography}[{\includegraphics[width=1in,height=1.25in,clip,keepaspectratio]{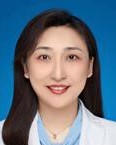}}]{Ping Wang}(Member, IEEE) received a B.S. degree in Automation from Shan Dong University, China, in 2004, an M.S. degree in Control Theory and Control Engineering from Shanghai Jiao Tong University, China, in 2007, and the PhD degree in Intelligent Robotics from Nanyang Technological University, Singapore, in 2011. She was a Professor Chang’an University, China. She is now an associate professor at the School of Intelligent Systems Engineering, Sun Yat-sen University, Guangzhou, China. Her current research interests include applications of control algorithms, artificial intelligence, and intelligent transportation systems.
\end{IEEEbiography}

\vspace{-5ex}

\begin{IEEEbiography}[{\includegraphics[width=1in,height=1.25in,clip,keepaspectratio]{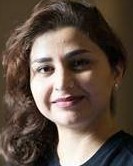}}]{Fatemeh Golpayegani}(Senior Member, IEEE) received the M.Sc. degree from Sharif Technical University and the Ph.D. degree in computer science from Trinity College Dublin. She is currently an Assistant Professor with the School of Computer Science, University College Dublin, where she leads the Multi-Agent Systems and Sustainable Solutions Research Group. Her research interests include intelligent agent-based models for dynamic and autonomous decision-making in large-scale and mobile environments.
\end{IEEEbiography}

\vspace{-5ex}

\begin{IEEEbiography}
[{\includegraphics[width=1in,height=1.25in,clip,keepaspectratio]{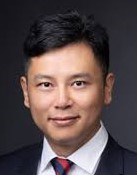}}]{Shen Wang}
(Senior Member, IEEE) received the M.Eng. degree from Wuhan University, China, and the Ph.D. degree from Dublin City University, Ireland. Some key industry partners of his applied research are IBM Research Brazil, Boeing Research and Technology Europe, and Huawei Ireland Research Centre. He is currently an Assistant Professor with the School of Computer Science, University College Dublin, Ireland. He has been involved with several EU projects as a co-PI, the WP, and a task leader of big trajectory data streaming for air traffic control and trustworthy AI for intelligent cybersecurity systems. His research interests include connected autonomous vehicles, explainable artificial intelligence, and security and privacy for mobile networks.

\end{IEEEbiography}

\end{document}